\documentclass[preprint,showpacs,preprintnumbers,amsmath,amssymb]{revtex4-1}
\usepackage{graphicx}
\usepackage{dcolumn}
\usepackage{bm}

\begin{document}

\title{Characterization of vector diffraction-free beams}

\author{Ting-Ting Wang,$^1$ Shuang-Yan Yang,$^1$ and Chun-Fang Li$^{1,2,}
\footnote{Email address: cfli@shu.edu.cn}$}

\affiliation{$^1$Department of Physics, Shanghai University, Shanghai 200444, China}

\affiliation{$^2$State Key Laboratory of Transient Optics and Photonics, Xi'an Institute
of Optics and Precision Mechanics of the Chinese Academy of Sciences, Xi'an 710119,
China}


\begin{abstract}

It is observed that a constant unit vector denoted by $\mathbf I$ is needed to
characterize a complete orthonormal set of vector diffraction-free beams. The previously
found diffraction-free beams are shown to be included as special cases. The $\mathbf
I$-dependence of the longitudinal component of diffraction-free beams is also discussed.

\end{abstract}

\maketitle


Any light beams other than plane waves are usually diffractively spreading in
propagation. But it was predicted \cite{Durnin} and then experimentally observed
\cite{Durnin-ME} that there exists a kind of $J_0$ ``scalar" mode the intensity of which
is free of diffraction. With the understanding \cite{Davis-P, Davis-P2, Jordan, Hall} of
the so-called cylindrical-vector beams \cite{Youngworth-B}, it was found \cite{Bouchal-O}
that there also exists a kind of cylindrical-vector modes the intensity and vectorial
structure of which is free of diffraction. Those two kinds of diffraction-free beams
share the same property that all the wavevectors of the constituent plane waves lie on
the surface of a cone.

The ``scalar" beam is in fact a uniformly polarized beam that is valid only in the
paraxial limit \cite{Lax-LM}. The cylindrical-vector beam is such a beam the direction of
whose electric vector is rotationally symmetric about its propagation axis. The problems
with which we are concerned here are whether there exist other kinds of diffraction-free
beams and whether we can find a precise scheme to distinguish between different
diffraction-free beams. Recently, a characteristic denoted by a constant unit vector
$\mathbf I$ was demonstrated \cite{Li} to convey the vectorial nature of a light beam.
The uniformly polarized beam has a characteristic vector that is perpendicular to the
propagation direction. The cylindrical-vector beam has a characteristic vector that is
parallel to the propagation direction. The purpose of this paper is to classify the
vector diffraction-free beams with the $\mathbf I$ and to explore the dependence of their
vectorial property on the $\mathbf I$. It will be shown that the $\mathbf I$ is a
continuous index to characterize a complete orthonormal set of vector diffraction-free
beams. Different $\mathbf I$'s represent different complete orthonormal sets of vector
diffraction-free beams.

For simplicity, only monochromatic light beams are considered. We do not solve the vector
Helmholtz equation together with the transversality condition. Instead, we directly make
use of the transversality condition to write out the integral expression for the
diffraction-free beams, because such an approach explicitly demonstrates \cite{Li} the
necessity of introducing the characteristic vector. As we know, the electric vector
$\mathbf{f}(\vartheta, \varphi)$ of a monochromatic light beam in the momentum
representation can be factorized \cite{Akhiezer} into a polarization vector
$\mathbf{e}(\vartheta, \varphi)$ and a scalar magnitude $f(\vartheta, \varphi)$ as
\begin{equation}\label{factorization}
    \mathbf{f}(\vartheta, \varphi) =\mathbf{e}(\vartheta, \varphi) f(\vartheta, \varphi),
\end{equation}
where $\vartheta$ and $\varphi$ are the polar and azimuthal angles of the wavevector
$\mathbf k$, respectively, in the spherical polar coordinates. The transversality
condition means that the polarization vector $\mathbf{e}$ is perpendicular to the
wavevector and can be expanded in terms of a set of base polarization vectors as
\begin{equation}\label{polarization vector}
    \mathbf{e}(\vartheta, \varphi; \mathbf{I})
        =\alpha_1 \mathbf{e}_1 (\vartheta, \varphi; \mathbf{I})
        +\alpha_2 \mathbf{e}_2 (\vartheta, \varphi; \mathbf{I}),
\end{equation}
where the base vectors $\mathbf{e}_1$ and $\mathbf{e}_2$ are defined \cite{Li} by means
of a constant unit vector $\mathbf I$ as
\begin{equation}\label{base polarization vector}
    \mathbf{e}_1 (\vartheta, \varphi; \mathbf{I})=\mathbf{e}_2 \times \frac{\mathbf k}{k},
         \hspace{5pt}
    \mathbf{e}_2 (\vartheta, \varphi; \mathbf{I})=\frac{\mathbf{k} \times \mathbf{I}}
         {|\mathbf{k} \times \mathbf{I}|},
\end{equation}
$\alpha_1$ and $\alpha_2$ are complex constants satisfying $|\alpha_1|^2 +|\alpha_2|^2
=1$, and the dependence of $\mathbf e$ on the $\mathbf I$ is explicitly shown. In
addition, the scalar magnitude $f(\vartheta, \varphi)$ for an arbitrary monochromatic
beam of wave number $k$ can be expanded in terms of the following complete orthonormal
set of scalar functions,
\begin{equation}\label{scalar base function}
    f_{k_z l}(\vartheta,\varphi)
   =\frac{\delta(\vartheta-\vartheta_0)}{i^l \sqrt{2 \pi k} \sin \vartheta_0}
    e^{i l \varphi}, \hspace{5pt} l=0, \pm1, \pm2...
\end{equation}
where the longitudinal component of the wavevector, $k_z = k \cos \vartheta_0$, is chosen
to be one of the indices. The orthonormality property assumes the form
\begin{equation}\label{orthogonal}
    \int f^{\ast}_{k'_z l'} f_{k_z l} \sin \vartheta d \vartheta d \varphi
   =\delta_{l'l} \delta(k'_z-k_z),
\end{equation}
where $k'_z = k \cos \vartheta'_0$. The electric vector of the beam in the position
representation that is associated with the momentum-representation electric vector
(\ref{factorization}) is given by
\begin{equation}\label{integral}
\mathbf{E}(\mathbf{I}) =\frac{1}{2 \pi} \int \mathbf{f} (\vartheta, \varphi; \mathbf{I})
e^{i \mathbf{k} \cdot \mathbf{x}} \sin \vartheta d \vartheta d \varphi.
\end{equation}
From the complete orthonormal set of base vectors (\ref{base polarization vector}) and
the complete orthonormal set of scalar functions (\ref{scalar base function}), one
readily writes down the following complete orthonormal set of vector functions,
\begin{equation}\label{vector base function}
    \mathbf{f}_{\sigma,k_z l} (\vartheta, \varphi; \mathbf{I}) =\mathbf{e}_{\sigma}
        (\vartheta, \varphi; \mathbf{I}) f_{k_z l} (\vartheta, \varphi),
\end{equation}
where $\sigma=1, 2$. They satisfy the relation
\begin{equation}\label{vector orthogonal}
    \int \mathbf{f}^{\ast}_{\sigma',k'_z l'} \cdot \mathbf{f}_{\sigma,k_z l}
    \sin \vartheta d \vartheta d \varphi =\delta_{\sigma' \sigma} \delta_{l' l}
    \delta(k'_z-k_z).
\end{equation}

Now we are in a position to show that the beams associated with the electric vector
(\ref{vector base function}) in the momentum representation are diffraction-free.
Substituting Eqs. (\ref{vector base function}) and (\ref{scalar base function}) into Eq.
(\ref{integral}) and performing the integration over $\vartheta$, one finds
\begin{equation}\label{diffraction-free beam}
    \mathbf{E}_{\sigma,k_z l} (\mathbf{I}) =\frac{e^{i k_z z}}{2 \pi i^l}
     \int \mathbf{e}_{\sigma} (\vartheta_0, \varphi; \mathbf{I})
     e^{il \varphi} e^{i \kappa \rho \cos(\phi-\varphi)} d\varphi,
\end{equation}
where $\kappa =k \sin \vartheta_0$ is the transverse component of the wavevector, the
position vector $\mathbf x$ is expressed as $\mathbf{x} =\mathbf{e}_x \rho \cos \phi
+\mathbf{e}_y \rho \sin \phi +\mathbf{e}_z z$ in the circular cylindrical coordinates,
and an irrelevant factor $1/\sqrt{2 \pi k}$ is omitted. The beams represented by Eq.
(\ref{diffraction-free beam}) are indeed diffraction-free, because only the propagation
factor $\exp(i k_z z)$ depends on the $z$ coordinate. Eqs. (\ref{vector base function})
and (\ref{scalar base function}) show that all the wavevectors in these diffraction-free
beams lie on the surface of a cone, the cone angle of which is $\vartheta_0$.

The index $\sigma$ in Eq. (\ref{vector base function}) as well as Eq.
(\ref{diffraction-free beam}) expresses the restriction imposed by the transversality
condition. Although it requires that the base vectors $\mathbf{e}_{\sigma}$ be
perpendicular to the wavevector, the transversality condition itself is not able to
prescribe \cite{Li} the exact relations of $\mathbf{e}_{\sigma}$ with the wavevector.
Those relations are determined here by the unit vector $\mathbf I$ in Eqs. (\ref{base
polarization vector}). This shows that one needs to use $\sigma$ as well as $\mathbf I$
together to characterize the vectorial nature of a vector diffraction-free beam. Since
every specified $\mathbf I$ defines a complete orthonormal set of vector diffraction-free
beams as is explicitly indicated in Eq. (\ref{diffraction-free beam}), the $\mathbf I$
turns out to be an index to characterize such a complete orthonormal set. It is noted
that the $\mathbf I$ is always perpendicular to $\mathbf{E}_{2,k_z l}$.

Next, let us show how the previously found diffraction-free beams can be obtained from
Eq. (\ref{diffraction-free beam}). To this end, we let $\mathbf I$ lie in the $xoz$
plane,
\begin{equation}\label{vector I}
    \mathbf{I} =\mathbf{e}_x \sin \Theta +\mathbf{e}_z \cos \Theta,
\end{equation}
paying our attention only to the effect of its polar angle $\Theta$. This is because the
angular-spectrum function (\ref{scalar base function}) is rotationally symmetric about
the $z$ axis. Rotation of $\mathbf I$ about the $z$ axis amounts to a rotation of the
diffraction-free beam in the same way. In the first place, we assume that the $\mathbf I$
is perpendicular to the $z$ axis, $\Theta= \frac{\pi}{2}$. In this case, one finds from
Eq. (\ref{base polarization vector})
\begin{subequations}
\begin{align}
  \mathbf{e}_1 =&  \mathbf{e}_x (1 -\sin^2 \vartheta \cos^2 \varphi)^{1/2} \nonumber \\
                &  -\frac{\mathbf{e}_y  \sin\vartheta \sin\varphi
                         +\mathbf{e}_z \cos \vartheta}
  {(1 -\sin^2 \vartheta \cos^2 \varphi)^{1/2}} \sin\vartheta \cos\varphi, \\
  \mathbf{e}_2 =&  \frac{\mathbf{e}_y \cos\vartheta -\mathbf{e}_z \sin\vartheta \sin\varphi}
                        {(1 -\sin^2 \vartheta \cos^2 \varphi)^{1/2}}.
\end{align}
\end{subequations}
Both of them have longitudinal components. Here $\sin \vartheta$ corresponds to the small
number $f$ discussed in Ref. \cite{Lax-LM} when the wavevector cone is very close to the
$z$ axis. To the zeroth-order paraxial approximation, one has $\mathbf{e}_1 \approx
\mathbf{e}_x$ and $\mathbf{e}_2 \approx \mathbf{e}_y$. Substituting them into Eq.
(\ref{diffraction-free beam}), one obtains
\begin{equation}\label{uniformly polarized}
    \mathbf{E}_{1,k_z l} =\mathbf{e}_x J_l e^{il \phi} e^{i k_z z},
    \hspace{5pt}
    \mathbf{E}_{2,k_z l} =\mathbf{e}_y J_l e^{il \phi} e^{i k_z z},
\end{equation}
where $J_l= J_l (\kappa \rho)$ is the $l$th-order Bessel function of the first kind. The
case of $l=0$ leads to the uniformly polarized $J_0$ diffraction-free beam found in Refs.
\cite{Durnin, Durnin-ME}. In the second place, we assume that the $\mathbf I$ is parallel
to the $z$ axis, $\Theta=0$. In this case, one has $\mathbf{e}_1 =-\mathbf{e}_{\kappa}
\cos \vartheta +\mathbf{e}_z \sin \vartheta$ and $\mathbf{e}_2 =-\mathbf{e}_{\varphi}$.
Upon substituting into Eq. (\ref{diffraction-free beam}), one gets
\begin{subequations}
\begin{align}
    \mathbf{E}_{1,k_z l}=& \frac{i}{2}
        [ (\mathbf{e}_{\rho} +i \mathbf{e}_{\phi}) J_{l-1}
         -(\mathbf{e}_{\rho} -i \mathbf{e}_{\phi}) J_{l+1}] \nonumber \\
                         & \times \cos \vartheta_0 e^{il \phi} e^{i k_z z}
         +\mathbf{e}_z J_{l} \sin \vartheta_0 e^{il \phi} e^{i k_z z}, \\
    \mathbf{E}_{2,k_z l}=& \frac{1}{2}
        [ (\mathbf{e}_{\rho} +i \mathbf{e}_{\phi}) J_{l-1}
         +(\mathbf{e}_{\rho} -i \mathbf{e}_{\phi}) J_{l+1}] \nonumber \\
                         & \times e^{il \phi} e^{i k_z z}.
\end{align}
\end{subequations}
Here $\mathbf{E}_{1,k_z l}$ and $\mathbf{E}_{2,k_z l}$ are exactly the vector solutions
$\mathbf{N}_n$ and $\mathbf{M}_n$, respectively, found in Ref. \cite{Bouchal-O}. In fact,
the authors of Ref. \cite{Bouchal-O} also made use of a constant vector. Unfortunately,
they just assumed that vector to be the unit vector in the $z$ direction.

In Eq. (\ref{diffraction-free beam}), we chose the linearly-polarized base vectors
$\mathbf{e}_{\sigma}$ for the vector diffraction-free beams. If we choose the
circularly-polarized base vectors, $\mathbf{e}_{\pm} =\frac{1}{\sqrt 2} (\mathbf{e}_1 \pm
i \mathbf{e}_2)$, and let the $\mathbf I$ lie along the $z$ axis, we will arrive at a
complete orthonormal set of vector diffraction-free beams that was found in Ref.
\cite{Enk-N}.

At last, let us have a look at the effect of $\mathbf I$ on the vectorial property of the
diffraction-free beams by examining the $\mathbf I$-dependence of their longitudinal
components. For this purpose, we take $l=2$ as an example. With the $\mathbf I$ being
given by Eq. (\ref{vector I}), we have for the $z$ components of $\mathbf{e}_1$ and
$\mathbf{e}_2$, respectively,
\begin{subequations}
\begin{align}
  e_1^z &=  \frac{\sin^2 \vartheta \cos\Theta -\sin\vartheta \cos\vartheta \cos\varphi \sin \Theta}
                {[1-(\sin\vartheta \cos\varphi \sin\Theta+\cos\vartheta \cos\Theta)^2]^{1/2}}, \\
  e_2^z &= -\frac{\sin \vartheta \sin \varphi \sin \Theta}
                {[1-(\sin\vartheta \cos\varphi \sin\Theta+\cos\vartheta
                \cos\Theta)^2]^{1/2}}.
\end{align}
\end{subequations}
The $z$ components of $\mathbf{E}_{\sigma, k_z 2}$ are thus given by
\begin{equation}\label{longitudinal component}
    E^z_{\sigma,k_z 2} =-\frac{e^{i k_z z}}{2 \pi} \int e_{\sigma 0}^z
     e^{2i \varphi} e^{i \kappa \rho \cos(\phi-\varphi)} d\varphi,
\end{equation}
where $e_{\sigma 0}^z =e_{\sigma}^z |_{\vartheta= \vartheta_0}$. The intensity of
longitudinal component is defined as
\begin{equation*}
    I^z_{\sigma, k_z 2}=|E^z_{\sigma, k_z 2}|^2.
\end{equation*}
Furthermore, in order to show the non-paraxial feature of the diffraction beam, the cone
angle of the wavevector cone is chosen to be $\vartheta_0= 60^{\circ}$.
\begin{figure}[ht]
\includegraphics[scale=0.6]{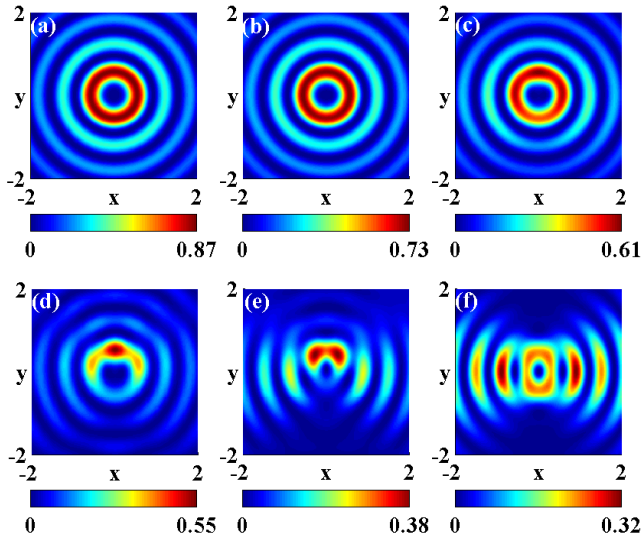}
\caption{Distributions of $I^z_{1,k_z 2}$ at a cross section for (a) $\Theta=0$, (b)
$\Theta =\frac{\pi}{10}$, (c) $\Theta =\frac{\pi}{5}$, (d) $\Theta =\frac{3 \pi}{10}$,
(e) $\Theta =\frac{2 \pi}{5}$, (f) $\Theta =\frac{\pi}{2}$. The units of $x$ and $y$ are
in wavelengths.} \label{fig1}
\end{figure}
In Fig. 1 are displayed the distributions of $I^z_{1,k_z 2}$ at a cross section for
different values of $\Theta$, where the units of $x$ and $y$ are in wavelengths. In order
to illustrate the relative strength of longitudinal component, $I^z_{1,k_z 2}$ is
normalized in each part by the maximum of the corresponding beam's intensity, $\max\{
|\mathbf{E}_{1,k_z 2}|^2 \}$. It is seen that with the increase of $\Theta$, the
longitudinal component of $\mathbf{E}_{1,k_z 2}$ goes weaker. Only when the $\mathbf I$
is parallel to the $z$ axis, is the intensity of longitudinal component axially
symmetric.
\begin{figure}[ht]
\includegraphics[scale=0.6]{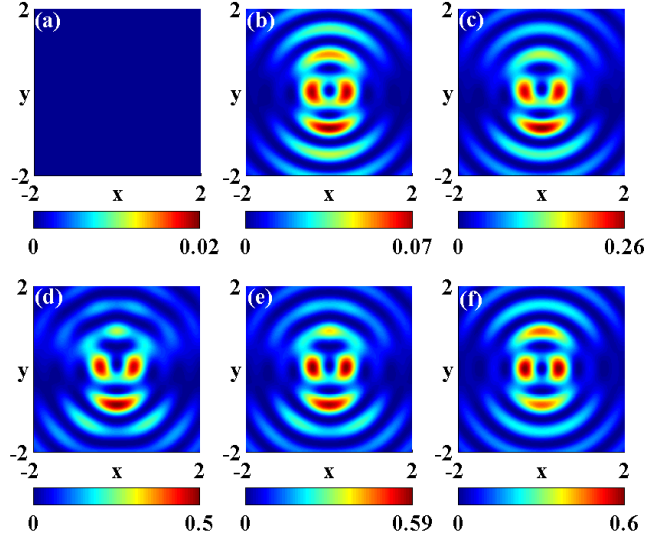}
\caption{Distributions of $I^z_{2, k_z 2}$ at a cross section for the same values of
$\Theta$ as in Fig. 1. The units of $x$ and $y$ are in wavelengths.} \label{fig2}
\end{figure}
For comparison, in Fig. 2 are displayed the distributions of $I^z_{2,k_z 2}$ at a cross
section for the same values of $\Theta$ as in Fig. 1, where the units of $x$ and $y$ are
in wavelengths, and the $I^z_{2,k_z 2}$ in each part is normalized as well by the maximum
of the corresponding beam's intensity, $\max \{ |\mathbf{E}_{2,k_z 2}|^2 \}$. It is shown
that with the increase of $\Theta$, the longitudinal component of $\mathbf{E}_{2,k_z 2}$
goes stronger. When the $\mathbf I$ is parallel to the $z$ axis, the longitudinal
component totally vanishes, in agreement with the fact that the $\mathbf I$ is
perpendicular to $\mathbf{E}_{2,k_z l}$.

It seems to be an accepted criterion \cite{Martinez} that whether a light beam can be
viewed as a paraxial beam depends on whether its longitudinal component can be neglected
in comparison with its transverse component. This should be valid when the characteristic
vector $\mathbf I$ is perpendicular to the propagation direction, because all the
diffraction-free beams in this case have negligible longitudinal components when the
wavevector cone is close to the $z$ axis, as is shown in Eqs. (\ref{uniformly
polarized}). But we have noticed that when the $\mathbf I$ is parallel to the propagation
direction, the beam associated with $\mathbf{E}_{2,k_z l}$ does not have longitudinal
component, regardless of the cone angle of wavevector cone. Because large cone angles
correspond to non-paraxial diffraction-free beams, the aforementioned criterion for a
beam to be paraxial is not strictly valid. Such a criterion needs further exploration.

This work was supported in part by the National Natural Science Foundation of China
(60877055 and 60806041) and the Shanghai Leading Academic Discipline Project (S30105).

\end{document}